\documentclass[11pt,twoside,openany]{article}
\usepackage{
	enumerate,
	amsmath,
	graphics,
	amssymb,
	graphicx,
	amscd,
	amsbsy,
	multirow,
	float,
	booktabs,
	verbatim,
	color,
	xcolor,
	amsfonts,
	latexsym,
	times,
	natbib,
	geometry}

\graphicspath{{figs/}{figs2/}{diagrams/}}
\newcommand{\E}{\mathop{{}\mathbb{E}}}
\newcommand{\var}{\text{Var}}

\newcommand{\bZ}{{\bf Z}}

\newcommand{\bbeta}{{\bf \beta}}

\newcommand{\R}{\mathbb{R}}

\newcommand{\QED}{\hspace*{\fill}\rule{2.5mm}{2.5mm}}

\newcommand{\labelone}{damped_with_lag_transformation}
\newcommand{\labeltwo}{damped_no_lag_transformation}
\newcommand{\labelthree}{undamped_with_lag_transformation}
\newcommand{\complexityone}{count-poisson-log-seas_growth}
\newcommand{\complexitytwo}{count-poisson-log-seas_growth_longt}
\newcommand{\complexitythree}{count-poisson-log-seas_growth_lag}
\newcommand{\complexityfour}{count-poisson-log-seas_growth_longt_lag}

\begin{document}

	\begin{center}
		{\large \bf Model selection for count timeseries with applications in\\ forecasting number of trips in bike-sharing systems and its volatility}\\
		\vspace{0.1cm}
		{Alireza Hosseini$^{1}$, Reza Hosseini$^{2}$}\\
		{\small Address$^{1}$: Yazd University, University Blvd, Safayieh, Yazd, Iran}\\
		{$^1$email: arh31415@gmail.com, $^2$email: reza1317@gmail.com}
	\end{center}

\begin{abstract}
	Forecasting the number of trips in bike-sharing systems and its volatility over time is crucial for
	planning and optimizing such systems.
	This paper develops timeseries models to forecast hourly count timeseries data,
	and estimate its volatility.
	Such models need to take into account the complex patterns
	over various temporal scales including hourly, daily, weekly and annual as
	well as the temporal correlation. To capture this complex structure, a large number
	of parameters are needed. Here a structural model selection approach is utilized
	to choose the parameters.
	This method explores the parameter space for a group of covariates at each step. These groups of covariate are constructed to represent a particular structure in the model.
	The statistical models utilized are extensions of Generalized Linear Models to
	timeseries data. One challenge in using such models is the explosive behavior of
	the simulated values.
	To address this issue, we develop a technique which relies on damping the
	simulated value, if it falls outside of an admissible interval.
	The admissible interval is defined using measures of variability of
	the left and right tails. A new definition of outliers is proposed based on
	these variability measures. This new definition is shown to be useful
	in the context of asymmetric distributions.
\end{abstract}

\section{Introduction}
Bike-sharing systems are being developed in many urban areas as a low cost technology
to solve mobility of millions of people. These systems have
access to rich data over time, since the timestamp of each check-in and check-out are collected automatically.
These data can be used to calculate the number of trips in any time-interval,
in a certain area or the number of trips between any two areas in the city.
The methods discussed in this paper apply to other ride-sharing systems using cars (e.g.\;Uber, Lyft)
and public transit.

The focus of this paper is to develop models to estimate the number of
trips at any given time. We also require the models to estimate the remaining volatility of
the series -- not explained by seasonality and long-term trends.
Some applications of these models are (a) short-term and long-term planning for the
demand in the system; (b) investigating the impact of interruptions in the system;
(c) planning for re-balancing of the bikes throughout the system;
(d) anomaly detection, e.g.\;for detecting events in the city and labeling them
as discussed in \cite{fanaee-2013}.

The contributions of this paper are as follows. From an application point of
view, most of the work in this area (ride-sharing) are focused on
forecasting the number of trips in short-term (in the order of a few days/weeks into the future)
using machine learning algorithms without an explicit model for volatility e.g.\;\cite{fanaee-2013} and \cite{yang-2016}.
Here, we provide long-term forecasts
and more importantly accurate estimates of the volatility of the series over time.

From a methodological point of view, a comprehensive summary of previous work
can be found in \cite{fokianos-2012}, which
classifies the methods into two categories:
(1) Regression Timeseries Methods (e.g.\;\cite{kedem});
(2) Integer Autoregressive Models (e.g\;\cite{nastic-2012}).
This work uses (1), because it can explicitly model
complex temporal variations in the series as shown in \cite{hosseini-takemura-2015}
for Gaussian series with time-varying variance or in \cite{hosseini-pcpn-2017}
for non-negative valued timeseries (e.g.\;precipitation).
The statistical timeseries models utilized in this paper are based on extensions of Generalized
Linear Models to timeseries context (e.g.\;\cite{kedem}, \cite{hosseini-takemura-2015}).
In order to apply these models, one challenge is the complexity of the temporal trends in various scales: daily, weekly and annual and the large number of lags needed
to capture the time-dependence. Modeling these features require a large number of parameters,
which can cause over-fitting and instability in the models.
The issue of over-fitting is extensively discussed in the machine learning context and
statistics (e.g.\;\cite{hastie-2009}). Various model selection techniques
have been developed to solve this problem. There are three main techniques to avoid
over-fitting
(1) cross-validation methods (see \cite{hastie-2009});
(2) methods which penalize the complexity of the model;
(e.g.\;AIC, \cite{akaike-1974} or BIC, \cite{schwartz-1978})
(3) regularization methods such as Ridge Regression and Lasso (see \cite{hastie-2009}).
The method utilized here is called {\it Grouped Structural Model Selection (GSMS)} and it falls under (2).
At each step of GSMS, rather than considering one variable, a group of variables -- which represent a particular
structure of the series -- is considered. This is the first application of this method to
count time-series data and at the hourly granularity.

While model selection helps with the stability of the model and allows us to
fit very complex patterns, an additional fundamental complexity needs to be addressed to ensure:
the simulated long-term future series do not diverge (produce unrealistic values, which are very far from historical values).
\cite{tong-1990} and \cite{hosseini-takemura-2015} discuss methods to avoid this
issue in the context of non-linear Gaussian series and in \cite{hosseini-pcpn-2017}
in the context of Bernouli-Gamma series for precipitation modeling.
For simplicity, assume $Y(t) = f(t) + \epsilon(t),$ where
$f(t)$ is a complex function of previous lags (e.g.\;includes $Y(t-1)$)
and $\epsilon(t)$ is a noise process. To avoid this potential explosive behavior,
\cite{tong-1990} suggested hard or soft censoring of $f(t)$ by replacing it with
$f(t)1_{(-a,a)}(f(t))$ which sets $f(t)$ to zero if it goes beyond the interval
$(-a,a)$ for some positive value of $a$. \cite{hosseini-takemura-2015} used a
damping approach by setting back $Y(t)$ to $M$ if $Y(t)>M$ and to $m$ if
$Y(t)<m$ for some appropriate real numbers $m<M$ chosen based on historical
observations.
However, for some data sets and models,
this is not sufficient and we suggest a method which lets $m$ and $M$ to depend
on some notion of seasonality. Moreover, in order to
find an upper bound ($M$) and lower bound ($m$), we present a new definition of outliers.
This definition allows for the lower and upper bounds
to use their own notion of variability depending on the left and right tails of the distribution
of the data.

The paper is organized as follows. Section \ref{sect:explore} performs an exploratory
analysis of the data. Section \ref{sect:models} discusses the underlying statistical models.
Section \ref{sect:model-selection} discusses the model selection procedure.
Section \ref{sect:damping} develops a damping procedure for the simulated values to insure
the simulated values do not diverge. This method allows for the damping parameters to depend on time.
To decide the upper bound and lower bounds for the damping, we introduce an asymmetric definition of
outliers which can be useful in many other applications.
Section \ref{sect:results} discusses the results of the model selection and damping procedure
for the bike-sharing data. Section \ref{sect:discussion} provides a summary and discussion
of the methods in this paper. Appendix \ref{sect:app-sims} includes comparison with a variety of
other related models to the main text.

\section{Exploratory analysis}
\label{sect:explore}
The dataset in this work includes the hourly counts of
rented bikes in Washington DC during 2011 and 2018 and were obtained from
{\it www.capitalbikeshare.com}.

\begin{figure}[H]
	\centering
	\includegraphics[scale=0.4]{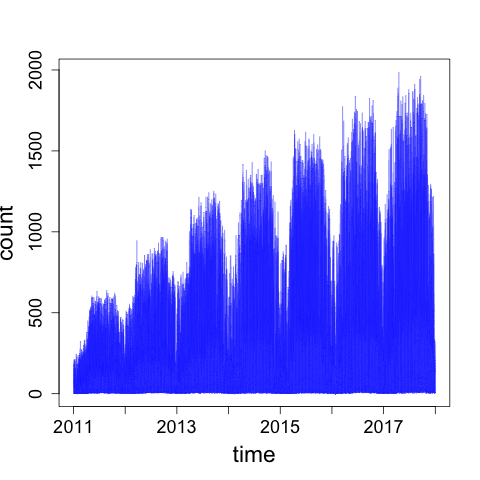}\includegraphics[scale=0.4]{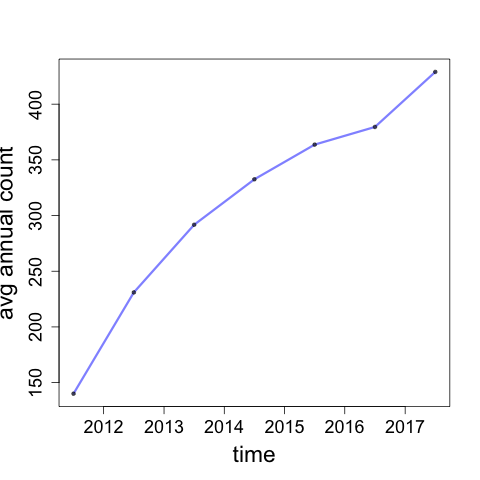}

	\includegraphics[scale=0.4]{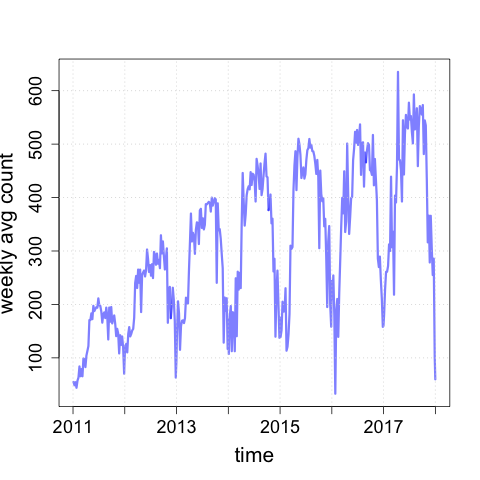}\includegraphics[scale=0.4]{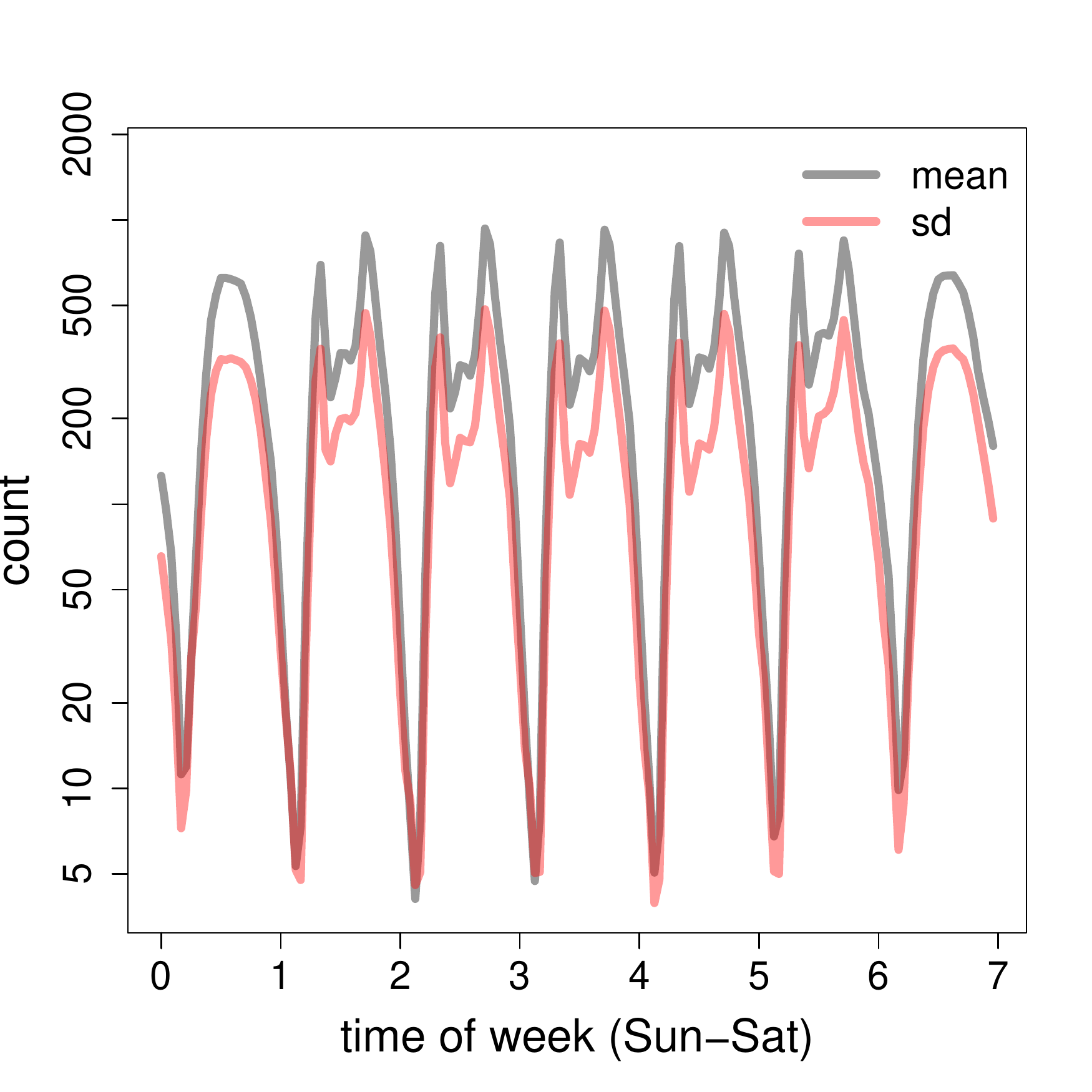}

	\caption{Top Left Panel: observed hourly count over time; Top Right Panel:
  annual average count (per hour); Bottom Left Panel: long-term weekly average
  count (per hour); Bottom Right Panel: observed mean and standard deviation with respect to the time of week.}
	\label{observed_count_over_time.png}
	\label{annual_avgs.png}
	\label{longterm_weekly_avg_count.png}
	\label{observed_data_mean_sd_var_with_respect_to_time_of_week.pdf}
\end{figure}

The raw hourly count data are shown in the Top Left Panel of Figure
\ref{observed_count_over_time.png}. The annual average count (per hour) curve is given
in the Top Right Panel, showing an increasing long-term trend. The Bottom Left Panel depicts the weekly average counts (per hour) and again showing a year long
seasonal pattern with largest counts during the warm season.
The Bottom Right Panel shows the average and the standard deviation of hourly count
with respect to the time of the week, starting from 0 (denoting Sunday 00:00 am), to 7 (denoting 24:00 on Saturday).
We observe a very strong weekly pattern in the average and variance measures.

\section{Statistical models}
\label{sect:models}
Suppose $\{Y(t)\}, t=0,1,\cdots$ is a count process, where $t$ denotes time.
We denote the available information up to time $t$ by $\mathcal{F}(t-1)$.
We assume this information is given in terms of a covariate process, denoted by
$\bZ(t)$, as discussed in \cite{kedem}.
For example, we can consider the covariate process:
\[\bZ(t-1) =  (1, Y(t-1), Y(t-2), f_{d}(t), f_{w}(t), f_{y}(t)),\]
where $,f_{d}(t), f_{w}(t), f_{y}(t)$ are each a vector of periodic functions over:
a day period; a week period; and a year period, respectively.

An example for $f_d$ is the following vector of terms of a Fourier series:
\[f_d(t) = \big ( \sin(\omega_d t),\cos(\omega_d t) \big ),\]
where $\omega_d = 2 \pi/24$ is the time frequency. This means the above
function is periodic with periodicity of 24, which is appropriate for hourly data.

We further assume, a linear form for the transformed conditional mean:
\begin{align}
	g(\E\{Y(t) | \mathcal{F}(t-1)\})&  = \bbeta \bZ(t-1),  \label{glm-eqn}
\end{align}
where $g$ is the transformation, also referred to as: the {\it link function}.

We assume, the conditional distribution of $Y(t)$ belongs to the Exponential
Family and has one of the following forms:
\begin{itemize}
	\item Poisson distribution, with density function:
	\[f_Y(y)={\frac {\lambda ^{y}}{y!}}e^{-\lambda },\;\; y=0,1,2,\cdots,\]
	where $\lambda = \E(Y) = \var(Y).$ In this case $\var(Y)/\E(Y)=1.$

	\item Negative Binomial distribution, with density function:
	\[\Pr(Y=y)={\frac {\Gamma (\theta+y)}{y!\,\Gamma (\theta)}}\left({\frac {\mu}{\theta+\mu}}\right)^{y}\left({\frac {\theta}{\theta+\mu}}\right)^{\theta}\quad {\text{for }}y=0,1,2,\dotsc \]
	where $\E(Y)=\mu$ and $\var(Y) = \mu + \mu^2/\theta.$ Therefore, we have
  $\var(Y)/\E(Y) = 1  + \mu/\theta$ which increases as $\mu$ increases.

\end{itemize}

For the Poisson distribution, the property: $\E(Y) = \var(Y),$ seems very limiting at first.
In fact, many authors report over-dispersion in real data,
in the context of Poisson regression (e.g.\;\cite{thall-1990}, \cite{berk-2008}
and \cite{zheng-2006}).
However, the above model is a conditional Poisson distribution of the form:
$\E\{Y(t) | \mathcal{F}(t-1)\})$. This means, the conditional
distribution, (potentially conditioned on a complex $\bZ(t-1)$), is distributed as a Poisson.
In fact, we are not assuming, the marginal distribution of
$Y(t)$ is Poisson. In particular, note that, if the covariate process: $\bZ(t-1)$, includes
complex seasonality components, growth components and lags (as
suggested above), some extra variability will be induced in the series, which might (or
might not) be sufficient to capture the true variability in the series.
In order to check for that, we compare the variability observed
in the model-based simulations to the observed variability on a validation set.

The Negative Binomial Model has an extra parameter, which allows the variability to
increase, as the conditional mean increases. Therefore, it can accommodate
more over-dispersion than the Poisson Model. In this work, we consider both Poisson and
Negative Binomial for a variety of
models with complex conditional means and compare the results.
It turns out that, with sufficiently complex conditional mean structures, even
the Poisson model is able to capture the time-varying volatility well.

Other related models considered in the literature include the Quasi-Poisson
Regression Model.
For example, \cite{verhoef-2007} compares the performance of Quasi-Poisson and
Negative Binomial
regression models to capture the over-dispersion. One issue with using Quasi-Poisson
models is the lack of a distribution
function to simulate from. One idea is to use the estimated
model's mean and
variance for the quasi-likelihood, and pick a distribution with matching mean and
variance (for example Gaussian, Gamma, or Poisson). We tested
this approach in this work, and the results were generally inferior to Negative
Binomial and Poisson. Therefore, we do not present them for brevity.
Another distribution proposed to capture
the over-dispersion
is the Double Poisson distribution introduced by \cite{efron-1986}. This distribution
is suggested to capture the over-dispersion in the timeseries context in \cite{heinen-2003}.

\section{Model selection}
\label{sect:model-selection}
Let $\{Y(t)\}, t=0,1,\cdots$, denote the hourly count data observed over time.
Then each time has these associated values:
\begin{itemize}
	 \item {\it time of day (shorthand: tod)} (in hours) which we denote by $d(t)$
	ranging from 0 to 23;
	\item {\it time of week (shorthand: tow)} ranging from 0 to 6 which we
	denote by $w(t)$ (with 0 denoting Sunday);
	\item {\it time of year (shorthand: toy)}
	which ranges from 0 to 365 (or 366) which we denote by $a(t)$.
\end{itemize}
 We allow these variables to change in the smallest scale available in the data. For example
 for noon of Monday Jan 15th we have: \[d(t)=12, w(t)=1.5, a(t)=14.5.\]

To capture the daily patterns, we use a Fourier terms of the form:
\[s_k(t) = \sin (k \omega_d d(t)),\;c_k(t) = \cos (k \omega_d d(t)), \;\omega_d = 2 \pi / 24,\]
\[k=1,\cdots,K_1\]
In order to allow for contrast between weekends and weekdays, we introduce
similar terms which are equal to the above over the weekends and zero otherwise:
\[\begin{cases} s_k'(t) = s_k(t),\;c_k'(t) = c_k(t), & \mbox{t in weekend} \\
s_k'(t)=0,\;c_k'(t)=0, & \mbox{otherwise} \end{cases}\]
\[k=1,\cdots,K_2\]

To allow extra day-to-day variation between weekdays, we introduce weekly Fourier series terms:
\[s_k^w(t) = \sin (k \omega_w w(t)),\;c_k^w(t) = \cos (k \omega_w w(t)), \;\omega_w = 2 \pi / 7.\]
\[k=1,\cdots,K_3\]
We also showed a seasonal effect throughout the year
(Figure \ref{longterm_weekly_avg_count.png}), which we capture by introducing
\[s_k^a(t) = \sin (k \omega_a a(t)),\;c_k^a = \cos (k \omega_a a(t)),\]
\[\omega_a = 2 \pi / 365,\; \omega_a = 2 \pi / 366\; (\mbox{for leap years}).\]
\[k=1,\cdots,K_4\]

To capture the temporal correlation, we consider the lags: $Y(t-1), Y(t-2), \cdots$.
To capture long-term dependence in the series, without using a large number of lags, we
consider the {\it long-term average lag processes:}
 \[av_j(t) = \sum_{i=1}^j Y(t-i)/j,\]
introduced in \cite{hosseini-bin-pcpn} and \cite{hosseini-bin-temp}
 for modeling precipitation and frost occurrence respectively.
For example, $av_{5} $ is the average of the past 5 time-periods (hours in for our data) and $av_{24}$
is the average amount over the past 24 hours.

One issue with using lags in the model, especially when the link function
($g$ in Equation \ref{glm-eqn}) is the $log$ function is, more chance of explosive behavior of the simulations.
One technique to decrease this chance is to use transformations of the lags as discussed in \cite{fokianos-2012}. Here, we observed
that using appropriate lag transformations indeed decreased the chance of explosive behaviors and used this lag transformation:
\[f(y(t-i)) = \log(0.1 + y(t-i)).\]
The constant 0.1 is added to prevent $f$ from being undefined at zero.
We recommend using a constant between (0.1, 1) here so that 0 is mapped to a value between (-1, 0).
Note that using a very small value e.g.\; 0.0001 maps 0 to -4 and the model might get stretched too far to
fit the 0 values, while in this context the difference between 0 counts and small counts is not significant in practice. In the main text, we present the results with lag transformations and in the appendix, we show the undesirable impact of removing the lag transformations from the model.

To capture the long-term trends/growth of the
series, as suggested by the Top Right Panel of Figure \ref{observed_count_over_time.png},
we use $t$ and its powers e.g.\;$t, t^{2}, t^{3}, t^{1/2}, t^{1/3}$ as covariate.
However, we only allow one of these variables in the model (as decided by the
model selection procedure), to prevent over-fitting when
simulating long-term future series.

Considering the large number of parameters needed to capture the variability
of the series over multiple temporal scales, an efficient model selection
procedure to reduce the number of parameters is needed. Here, we utilize the
{\it Grouped Structural Model Selection} (GSMS) approach proposed in
\cite{hosseini-RS-2015} and \cite{hosseini-takemura-2015} for estimating soil
moisture and daily temperature respectively using Gaussian models.
\cite{hosseini-pcpn-2017} utilized this approach for modeling daily
precipitation process using a conditional Bernoulli-Gamma for the occurrence
and amount. This paper is the first application of this method to hourly time
series data. Also this is the first application of GSMS to count data.

GSMS works by grouping the related covariates into groups, each representing a
particular structure of the model, e.g.\;daily patterns, weekly patterns, lags etc.
Then GSMS limits the search to the subsets of a group (a pre-defined set of covariates)
at each stage.

Table \ref{table-groups}
includes the proposed groups for this application with their corresponding elements (covariates).
The groups $tod, tod_{wd}, tow, toy$ are introduced to capture the trends in the
series over various temporal scales. The groups $lags$, and $avglag$ are introduced
to capture the remaining auto-correlation in the series after removing the
seasonal patterns and finally the $growth$ group is there to model the long-term
trends of the series over time (as a result of system expansion and population increase).
\begin{table}[ht]
	\centering
	\caption{Covariate groups considered for GSMS procedure are given in the table.
		Each group represents a structure in the model. For example $tod$ captures
		the hourly variation across the day.}
	\begin{tabular}{| l | l | l |}
		\hline
		Group (acronym) & Elements/Variables \\
		\hline
		time of day ($tod$) & $s_1,c_1,\cdots,s_{10},c_{10}$   \\
		\hline
		time of day in weekend  ($tod_{wd}$) &  $s_1',c_1',\cdots,s_{10}',c_{10}$' \\
		\hline
		time of week ($tow$) &  $s_1^w,c_1^w,\cdots,s_{10}^w,c_{10}^w$  \\
		\hline
		time of year ($toy$) & $s_1^a,c_1^a,\cdots,s_{10}^a,c_{10}^a$  \\
		\hline
		lags ($lags$) & $y(t-1),\cdots,y(t-10)$\\
		\hline
		long-term averaged lags ($avglag$) & $av_i,\; i=5,10,15,24,48$\\
		\hline
		long-term growth ($growth$) & $t, t^2, t^3, \sqrt{t}, \sqrt[3]{t}$\\
		\hline
	\end{tabular}
	\label{table-groups}
\end{table}
Here, we briefly describe the {\it Grouped Structural Model Selection} (GSMS)
method. Suppose, the predictors $S = \{z_1,\cdots,z_p\}$ are chosen already; $S'$ is
another set (group) of predictors; and $crit$ is a criterion to compare models
(e.g.\;AIC or BIC). We denote the value of the criterion for a set of predictors
$S$ by $crit(S)$. The union of two sets of predictors $S,S'$ is denoted by $(S \cup S')$.
Then, we define the {\it optimal merge}, $om,$ of $S'$ conditioned on the
predictors $S$, as follows:
\[om(S'|S,crit) = \underset{H \subset S'}{argmin}\; crit\{(H \cup S)\}.\]
In other words, $om$ picks the subset of $S'$ which once added to $S$ attains
the best performance according to the criterion $crit$. We use a sequential
approach combined with merging for the model selection performed in this work.

Other than the grouping, we also need to specify a {\it model selection diagram (flow)}
to describe how the algorithm scans through various groups. Note that mathematically the diagram is a
directed graph with nodes consisting of a group of covariates to optimize and the inward edges deliver the
optimal elements calculated in previous nodes.

\begin{figure}[H]
	\centering
	\includegraphics[scale=0.5]{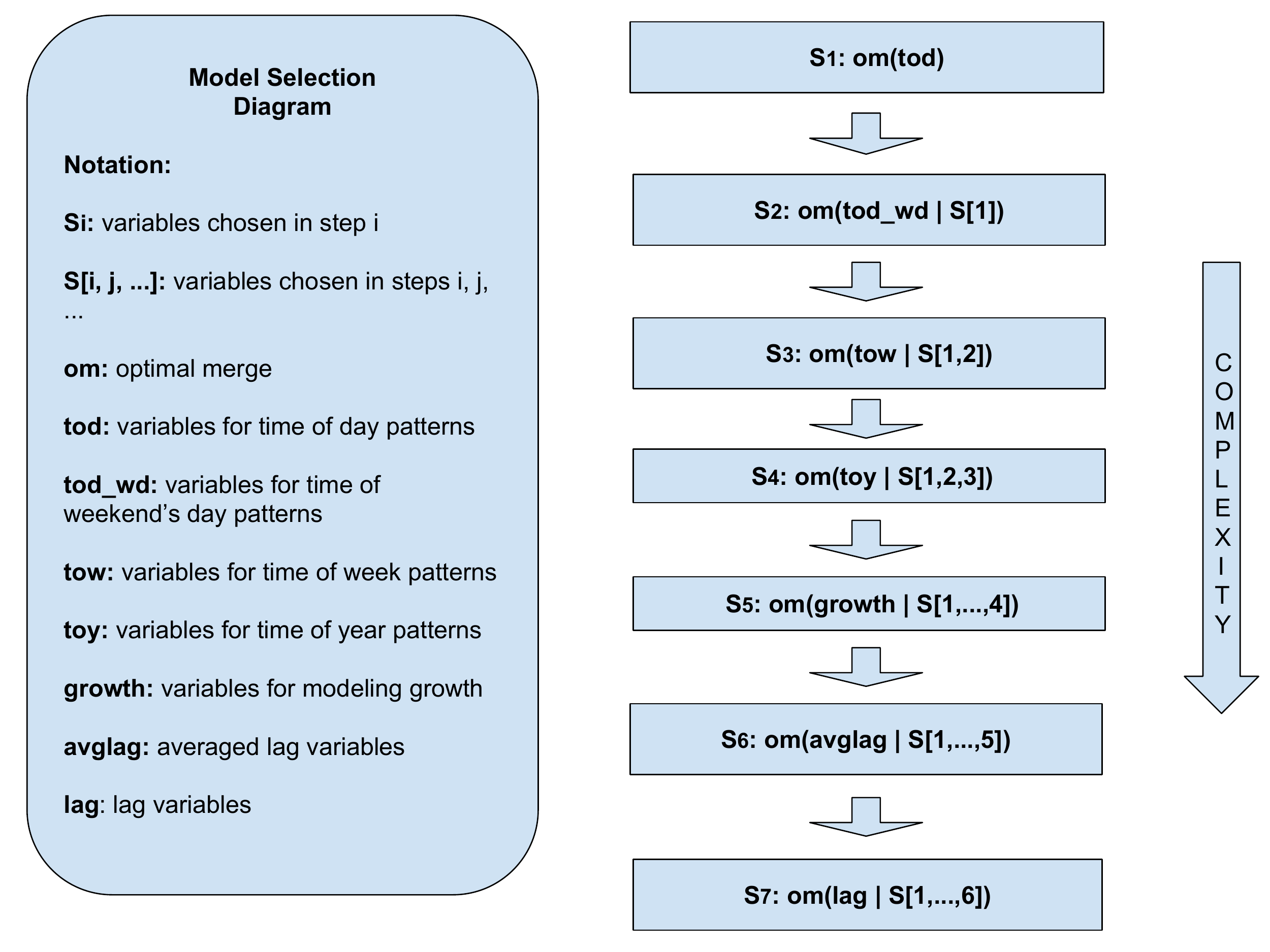}
	\caption{Model selection diagram for the bike-sharing data. In the first step ($S_1$),
		GSMS performs a search in the
		Fourier terms based on time of day ($tod$) and it proceeds to add patterns at other
		temporal scales in $S_2, S_3, S_4$.
		Then in $S_5$ it adds the long-term growth and finally in $S_6, S_7$ it allows for
		unexplained autocorrelation using lags and their averages.}
	\label{model_selection_count_process.pdf}
\end{figure}

Figure \ref{model_selection_count_process.pdf} shows the main model diagram in this paper, which is a fairly simple diagram.
See \cite{hosseini-takemura-2015}, \cite{hosseini-pcpn-2017} for more complex
GSMS diagrams.
In the following, we also consider other modeling diagrams which are simplifications
of the main one in
Figure \ref{model_selection_count_process.pdf}.
These simplifications are described in Table \ref{tab-other-model-diagrams}.

The advantages of using GSMS over some other model selection approaches such as
stepwise regression (see e.g.\;\cite{hastie-2009}) are:
\begin{itemize}
	\item[(1)] GSMS allows us to divide the covariates into groups of related covariates.
	Each group can represent a structural component of the model and we can assess the
	effect of adding or omitting a structure to the model rather than a single covariate.
	\item[(2)] We can use: (a) exploratory analysis results; (b) expert knowledge of the process; (3) and
	the robustness of the covariates; to decide which structures should be added to
	the model first. For example, from both expert knowledge and exploratory analysis,
	we know that daily patterns (captured by Fourier terms based on $tod$) can explain
	a large portion of the variability in the response and therefore we can add that
	structure to the model first. Also daily patterns as compared to lag terms have
	the advantage of being completely determined in the future and therefore are
	more robust. Note that, lag terms need to be filled in by simulated values when
	we are performing long-term forecasting. This can cause the series to start
	diverging to points in the input space which have not been observed, thus resulting
	in the model producing unreasonable values.
	\item[(3)] The GSMS approach allows us to
	fix a chosen component into the model (for example, the seasonal component) and
	then assess if adding another structure is beneficial.
\end{itemize}

\begin{table}[ht]
	\centering
	\begin{tabular}{|l|l|}
		\hline
		 model complexity string & structure (groups) \\
		\hline
		  seas\_only &   $(tod, tod_{wd}, tow, toy)$ \\
		  seas\_growth &  seas\_only + $growth$ \\
		  seas\_growth\_avglag & seas\_only + $avglag$ + $growth$   \\
		  seas\_growth\_lag &  seas\_only + $lag$ + $growth$ \\
		  seas\_growth\_avglag\_lag &  seas\_only + $avglag$ + $lag$ + $growth$ \\
		  \hline
	\end{tabular}
\caption{Various model complexity scenarios which are compared. In the first row,
	seas\_only is a model which only uses
	deterministic temporal patterns (no lags).
	Other rows add other structures to  seas\_only e.g.\;lags.}
\label{tab-other-model-diagrams}
\end{table}

\section{Time-varying damping}
\label{sect:damping}
In order to forecast the future values of the series and the volatility, we
simulate the future values using the fitted models. One issue with simulation
in the timeseries context is the possibility of explosive behavior for complex
models. This means, the simulated series is chaotic and does not look like the
observed series -- values far beyond what is been observed.
This is generally not an issue for standard predictive
models e.g.\;in an interpolation or classification problem. The reason
for the explosive behavior in timeseries context is two fold: (1) We are simulating values as opposed
to simply finding a good prediction which is the case in many machine learning
application. (2) We often need to simulate a much longer period than one time ahead
and as we move more into the future, the simulated values will depend on other
simulated values (e.g.\;due to presence of lags in the model). Hence, there
is more chance of gradually moving into a point in the input space (covariate space)
which is not seen before (by the trained model) and this could lead in simulating unrealistic values.

The explosive behavior in simulated timeseries values was discussed by
\cite{tong-1990} and \cite{hosseini-takemura-2015} for non-linear autoregressive
Gaussian processes. In that context, we have $Y(t) = f(t) + \epsilon(t)$ where
$f(t)$ is a polynomial function of previous lags and $\epsilon(t)$ is Gaussian
noise with potentially time-varying variance. To avoid the
explosive behavior, \cite{tong-1990} suggested hard or soft censoring
of $f(t)$ by replacing it with $f(t)1_{(-a,a)}(f(t))$ which sets $f(t)$ to zero
if it goes beyond the interval $(-a,a)$ for some positive value of $a$.
\cite{hosseini-takemura-2015} used the censoring idea for the $Y(t)$ process by
setting $Y(t)$ to
\[
	Y^{\star}(t) =
	\begin{cases}
        M, & Y(t) \geq M \\
        Y(t), & m < Y(t) < M \\
        m,  &  Y(t) \leq m
    \end{cases}
\]
for some real numbers $m < M$, to also
assure the noise process $\epsilon(t)$ does not become too large and cause
explosive behavior. In this approach, $m$ and $M$ can be chosen more naturally
using the historical observations.

Here, we extend the above method in two ways: (1) by allowing $(m, M)$ to vary with respect to
other variables; (2) by defining a more systematic way to pick $(m, M)$ which is more
reasonable for skewed and asymmetric distributions. For the bike-sharing application,
we allow $m, M$ to depend on the time of the week. This is because the response variable's
volatility shows a large variability with respect to the time of the week
(Figure \ref{observed_count_over_time.png}).

Since we know counts cannot be negative, and we do observe small values in the data, close to zero,
it is reasonable to assume $m=0$.
However, in order to illustrate how the method works in general,
we do not assume $m=0$.

Here, we describe a new method to define outliers, which considers the possibility of the
distributions being assymetric. There are several definition of outliers in the literature, some going back to the 19th century. As an example, \cite{chauvenet-1891} provides a definition using the mean
and standard deviation of the data, based on Gaussian distribution.
One of the most popular definitions of outliers is developed by John Tukey and discussed in \cite{tukey-1977}.
Tukey defines a point to be an outlier, if it falls outside the interval:
\[Q_{1}-k(Q_{3}-Q_{1}),Q_{3}+k(Q_{3}-Q_{1}),\]
where $Q_1, Q_3$ are the first and thrid quartiles of the data distribution and
$k$ is a constant. Tukey proposed to use $k=1.5$. The difference $\delta = Q_{3}-Q_{1}$ is
called the {\it Interquantile Range (IQR)}. The IQR is in utilized widely in statistics  going back to early 20th century (e.g.\;\cite{yule-1911} and \cite{upton-1996}). Note that the IQR measures
the variability of the data in the center of the distribution -- 50 percent of the data
is within that range.

In order to define the outliers, first we define new quantile-based measures of data variability.
The intuitive idea behind the following definitions is to measure the variability of the data
on the left and right side of the distribution instead of the center -- which is the case for
IQR.

\begin{figure}[H]
	\centering
	\includegraphics[width=15cm, height=20cm,keepaspectratio]{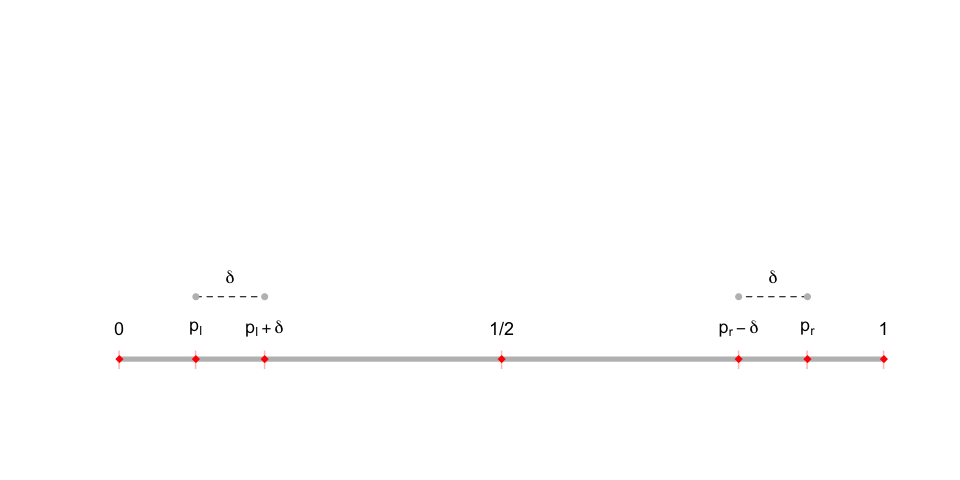}
	\caption{A demonstration of the values used in the definition of right and left tail variability given in Equation \ref{eqn-tail-var}.}
	\label{tail_var_definition.png}
\end{figure}

To that end, denote the distribution of the observed values by $D$ and
let $q_D(p)$ denote the quantile of $D$ for $p \in [0, 1]$. Note that, with this notation, the
IQR is given by
\[IQR(D) = q_D(0.75) - q_D(0.25).\]
Then consider three real numbers,
\[0< p_{r}, p_{l}, \delta_p < 1,\]
such that
\[0< p_{r}  < p_{r} + \delta_p <   0.5  < p_{l} - \delta_p < p_{l} < 0,\]
and require
\[p_l = 1 - p_r.\]
The indices $r$ and $l$ denote right and left respectively.
As an example, consider
\[p_r=0.01,\; p_{r} + \delta_p=0.1,\; p_{l} - \delta_p=0.9,\; p_l=0.99.\]
See Figure \ref{tail_var_definition.png} for a demonstration of these values
and their relationship.
Next, we define two measures of dispersion for the right and left tails to be
\begin{eqnarray*}
	RTV(D) &= \frac{q_D( p_{r} + \delta_p) - q_D(p_r)}{\delta_p},\\
	LTV(D) & = \frac{q_D(p_l) - q_D( p_{l} - \delta_p)}{\delta_p},
	\label{eqn-tail-var}
\end{eqnarray*}
and we call them right and left {\it tail-variability} ($TV$) respectively.
$RTV(D)$ and $LTV(D)$ can be interpreted as rate of change of quantile
function in $[p_r, p_r + \delta_p]$, and $[p_l - \delta, p_l]$ intervals respectively.

The difference between $TV$ and $IQR$ is: while $IQR$ measures the variation of data
in the center of the distribution, $RTV$ and $LTV$ measure the variation of the data on the
right and left sides/tails of the distribution.

Now we define the lower ($m$) and upper ($M$) limits for defining outliers:
\begin{align*}
	m &= q_D(p_m) - \alpha RTV(D)& \\
	M &= q_D(p_M) + \alpha LTV(D),& \\
	&\mbox{where } 0 < p_m \leq 1/2 \leq p_M,&\\
	&\mbox{and } p_M = 1 - p_m, \alpha \geq 0.&
\end{align*}
In order to contrast with Tukey's definition, note that:
 $q_D(p_m)$, and $q_D(p_M)$ play the role of $Q_1$ (First Quartile)
and $Q3$ (Third Quartile).
Here, we further assume that $p_m = p_r$ and $p_M = p_l$ for simplicity.

We can interpret these quantities as an extension to classical definition of
outliers which is useful in dealing with asymmetric distributions.
In order to find a rule of thumb to choose $\alpha$, note that
if hypothetically the true quantile function change approximately in the same
rate as $RTV/LTV$, the minimum/maximum value attained
by the distribution would be approximated by
\begin{equation} \label{eqn:approx-min-max}
  \begin{aligned}
 	q_D(0) & \approx q_D(p_m) - p_m RTV(D),& \\
 	q_D(1) & \approx q_D(p_M) + p_m LTV(D).& \\
	\end{aligned}
\end{equation}
Clearly, one should not expect the rate of change of the quantile function to remain
constant. However this calculation gives some idea of what can be a good rule of thumb.
For example if we assume the rate of change could be 10 times higher than above,
we can choose $\alpha = 10 p_m$.

Note that, for a continuous distribution such as Gaussian (or other distributions with positive-valued continuous density across $\R$)
$q_D(0) = -\infty$ and $q_D(1) = \infty$. Therefore, the above equations in (\ref{eqn:approx-min-max}), cannot hold.
However, one can argue that in most real world applications, in fact almost no variable can
have positive probability beyond a bounded interval including in our case. For example while we
might assume a Poisson conditional distribution, we know that actually a value beyond $10^9$ bikes
at any given time is impossible in one city for one hour.

\begin{figure}[H]
	\centering
	\includegraphics[width=0.9\textwidth]{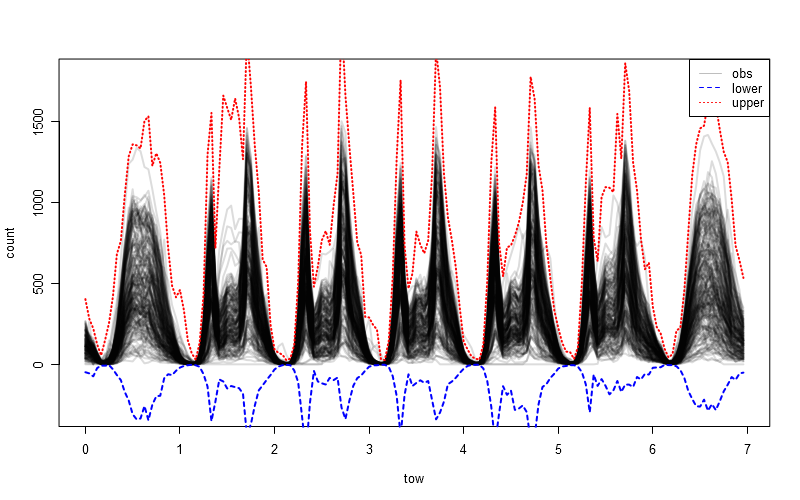}
	\caption{The obtained time-varying admissible intervals for for simulated are
		between the lower (denoting $m$, dashed curve) and upper bounds
		(denoting $M$, dotted curve) in the plot.
		The observed counts during 2011-2015 (training period) are also overlay-ed
		in the plot in transparent gray.}
	\label{derived_bounds.png}
\end{figure}

For this application we choose $p_r = p_m = 0.025$ an $alpha = 10 \times p_m = 2.5$.
Then we apply this to the observed distribution of the data for every hour of the week.
The results are given in Figure \ref{derived_bounds.png}, where the bounds $m, M$
depend on the time of the week and the obtained bounds are overlay-ed on top of
the observations for comparison. We can observe that the lower
bounds are closer to the historical observations since there is less variability
in the low values. As discussed before, in this case we have the external information that
the values can attain zero and cannot be negative, therefore a
reasonable choice is $m=0$ for all $t$.

We refer to the above procedure as the Time-Varying Damping Method ($TDM$).
It should be noted that, in general, models
which do not require damping are preferred. In fact, based on our experience, as a rule of thumb, if a model requires more than
20\% damping, one should be cautious about using the model and try to find an alternative model.
In other words, $TDM$ is useful when a model generally performs well but has some chance of drifting
into unrealistic values.

\section{Results}
\label{sect:results}
The results of the model selection and simulations for models with transformed lags
and with application of $TVM$ (Time-varying Damping Method) are given in
Tables \ref{\labelone_mod_info.tex}
and \ref{\labelone_sim_info.tex}.

Table \ref{\labelone_mod_info.tex}
contains the performance of the fit for various model complexity levels.
We have standardized the AIC and BIC values by dividing them by the number of observations.
In general, the models based on Negative Binomial (denoted by NB) have smaller AIC and BIC but the
correlation between observed and fitted are similar across model complexity for Negative Binomial and Poisson.

The seasonal only (seas\_only) model (which includes daily, weekly and annual Fourier terms) has
captured 84\% of the correlation already.
Also as we increase the complexity, the correlation increases with non-negligible jumps
when adding the growth components and either of the lag based components. However
adding the groups $lags$, $avglag$ (see Table \ref{table-groups} for definition)
or both resulted in comparable correlations.

In order to test the model performance further, we split the data to a training set
which includes the observations from beginning of 2011 to end of 2015 and a test set from
the beginning of 2016 to the end of 2018. The simulation results for each of the models
is given in Table \ref{\labelone_sim_info.tex}. This table reports: (i) the percent
of damped simulated values; (ii) the correlation between the
simulated values and the test set; (iii) the Mean Absolute Error (MAE);  and
(iv) Root Mean Square Error (RMSE).
We observe that in general the Poisson model
requires less damping as compared with the Negative Binomial model. Moreover, in general,
more complex models require more damping. The Poisson model with seasonal
components, growth component and $avglag$ has performed the best in terms of
MAE and RMSE and could be considered as the optimal model.

\begin{figure}[H]
	\centering
	\includegraphics[width=1.05\textwidth]{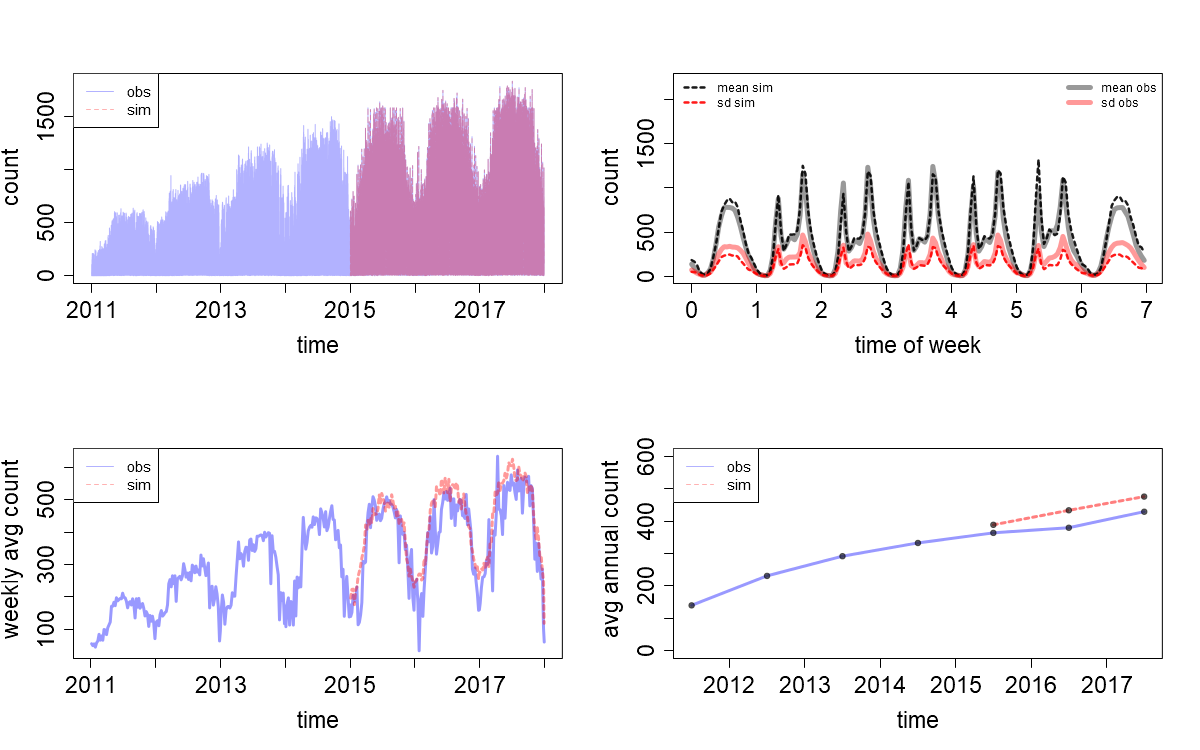}
	\caption{
		The simulation results from a Poisson conditional model with transformed
		lags and time-varying damping. The model is fitted on the data from 2011 to end of 2015
		and the validation data is considered to be from 2016 to end of 2018. The Top Left Panel shows the
		result of one simulation added to the observed data. The Top Right Panel shows the average and
		standard deviation of counts for each hour of the week from the observed data (on the validation set)
		and the model simulations, showing a good agreement. The Bottom Left
		Panel shows the average hourly count for each week of observed data and compare that to the simulated values.
		The bottom Left Panel shows the average annual count from the observed data and compares it
		to the average annual count from simulated data.}
	\label{\labelone_\complexityfour_sim_results.png}
\end{figure}

To visualize the performance of
the model, Figure \ref{\labelone_\complexityfour_sim_results.png} compares the
simulated values from a single simulation of the model with the test set. The Top Left Panel
shows the hourly simulations from the model in the test period and the simulations seem consistent with
the observations. The Top Right Panel compares the mean and the standard deviation of the observations
for each hour of the week with the simulated values, showing the model has been
able to capture complex weekly patterns in both the mean and volatility of the series.
In general, the
volatility seems to be captured well, despite
the model being a conditional Poisson.
The Bottom Left Panel compares the weekly averages of the observed data to
weekly averages of the simulated data. It shows a good agreement.
However, the simulated values are slightly higher overall. This fact is more clear from the
Bottom Right Panel which compares the annual averages.
Note that despite the large amount of data and the complexity of the model,
it is hard to forecast
long-term trends in growth because of the impact of many external factors which we cannot
infer about using the data. Such factors include population growth, policy changes and new investment.

\begin{table}[ht]
\centering
\begin{tabular}{|l|l|r|r|r|}
  \hline
Family & Parameter Complexity & std AIC & std BIC & cor(obs, fitted) \\
  \hline
Poisson & seas\_only & 57.80 & 57.80 & 0.84 \\
   \hline
Poisson & seas\_growth & 35.50 & 35.50 & 0.92 \\
   \hline
Poisson & seas\_growth\_longt & 20.00 & 20.10 & 0.97 \\
   \hline
Poisson & seas\_growth\_lag & 14.50 & 14.50 & 0.98 \\
   \hline
Poisson & seas\_growth\_longt\_lag & 14.40 & 14.40 & 0.98 \\
   \hline
NB & seas\_only & 11.20 & 11.20 & 0.83 \\
   \hline
NB & seas\_growth & 10.80 & 10.80 & 0.92 \\
   \hline
NB & seas\_growth\_longt & 10.20 & 10.20 & 0.96 \\
   \hline
NB & seas\_growth\_lag & 9.81 & 9.82 & 0.98 \\
   \hline
NB & seas\_growth\_longt\_lag & 9.78 & 9.79 & 0.98 \\
   \hline
\end{tabular}
\caption{Model fit results for models with lag transformation}
\label{damped_with_lag_transformation_mod_info.tex}
\end{table}

\begin{table}[ht]
\centering
\begin{tabular}{|l|l|r|r|r|r|}
  \hline
Family & Parameter Complexity & Percent Damping & RMSE & MAE & cor(obs, sim) \\
  \hline
Poisson & seas\_only & 9.75 & 251.00 & 167.00 & 0.90 \\
   \hline
Poisson & seas\_growth & 12.20 & 177.00 & 109.00 & 0.91 \\
   \hline
Poisson & seas\_growth\_longt & 12.30 & 173.00 & 106.00 & 0.91 \\
   \hline
Poisson & seas\_growth\_lag & 12.70 & 174.00 & 108.00 & 0.90 \\
   \hline
Poisson & seas\_growth\_longt\_lag & 12.30 & 171.00 & 106.00 & 0.90 \\
   \hline
NB & seas\_only & 10.40 & 261.00 & 170.00 & 0.86 \\
   \hline
NB & seas\_growth & 16.60 & 181.00 & 114.00 & 0.88 \\
   \hline
NB & seas\_growth\_longt & 14.10 & 177.00 & 111.00 & 0.88 \\
   \hline
NB & seas\_growth\_lag & 13.80 & 174.00 & 108.00 & 0.89 \\
   \hline
NB & seas\_growth\_longt\_lag & 12.90 & 176.00 & 110.00 & 0.89 \\
   \hline
\end{tabular}
\caption{Simulation results for damped models with lag transformation}
\label{damped_with_lag_transformation_sim_info.tex}
\end{table}

\section{Discussion}
\label{sect:discussion}
This paper developed models for count timeseries data which are able to capture
complex trends over time in the mean and volatility of the process.
We showed that by allowing
complex dependence on various temporal scales and lags, even conditional
Poisson models can achieve
the volatility observed in the data.

In order to deal with the large number of parameters needed,
we use the Grouped Structural Model Selection (GSMS) approach which
works by grouping the covariates into various groups, each of which represent
a structure in the model and then selecting the necessary covariates within
each group in each step.

One challenge which is common in many timeseries
applications is the possibility of explosive behavior in the simulations
from the model.
This problem is more likely, when there
is significant non-linearity in the conditional mean and the $log$ is the link
function. We tried some other link functions such as identity function, inverse and
square root. However, in those cases, the models often did not converge -- even
after providing careful initial values for the parameters
(e.g.\;by regressing the observed historical conditional probabilities on the covariates
as discussed in \cite{hosseini-bin-temp}). Moreover, while it is less likely to encounter
explosive behavior with these link functions, we still observed explosive behavior in some cases.
Based on these observations, we used the natural logarithm link function and to deal with the
explosive behavior, we developed a method which works by damping the simulated response when it
falls outside admissible intervals, which are defined using the observed data. To define
the admissible intervals, we first defined some quantile-based measures of variability
in the right and left sides/tails of the distribution. Using these variability measures,
we developed a new definition for outliers which works better for asymmetric distributions.
To improve flagging of outliers, we allowed the admissible
interval to depend on other variables (in our data on the hour of week).

\appendix

\section{Other simulations}
\label{sect:app-sims}
In this section, we provide the results for various other models which
do not use damping or lag transformation. In summary, the results
indicate that damping and lag transformation (with $f(x) = \log(x + 0.1)$
as the transformation) are beneficial in getting realistic simulations
from the models.

\subsection{Simulation with damping and no lag transformation}
In this subsection, we apply the models with no lag transformation
as opposed to using the log transform: $f(x) = \log(x + 0.1)$.

As we discussed in the main text,
the purpose of using transformed lag variables is to decrease the chance of explosive
behavior of the simulated values. This method is also suggested in \cite{fokianos-2012}.
We still continue to perform damping in this case, and the damping percentage can
measure how much explosive behavior is observed as a result of no lag transformations.
The results for the model fits are given in Table \ref{\labeltwo_mod_info.tex},
which are comparable with the model fits with lag transformations. However, Table
\ref{\labeltwo_sim_info.tex}, shows that for the models which involve lags, a much higher percent
of damping -- more than 50\% in some cases -- is needed. This indicates that, while
damping has kept the model simulations relatively realistic, the percentage of
damping is very high -- which questions the validity of the model in practice.

\begin{table}[ht]
\centering
\begin{tabular}{|l|l|r|r|r|}
  \hline
Family & Parameter Complexity & std\;AIC & std\;BIC & cor(obs, fitted) \\
  \hline
Poisson & seas\_only & 57.80 & 57.80 & 0.84 \\
   \hline
Poisson & seas\_growth & 35.50 & 35.50 & 0.92 \\
   \hline
Poisson & seas\_growth\_longt & 26.90 & 26.90 & 0.95 \\
   \hline
Poisson & seas\_growth\_lag & 25.40 & 25.40 & 0.95 \\
   \hline
Poisson & seas\_growth\_longt\_lag & 24.60 & 24.70 & 0.95 \\
   \hline
NB & seas\_only & 11.20 & 11.20 & 0.83 \\
   \hline
NB & seas\_growth & 10.80 & 10.80 & 0.92 \\
   \hline
NB & seas\_growth\_longt & 10.60 & 10.60 & 0.93 \\
   \hline
NB & seas\_growth\_lag & 10.60 & 10.60 & 0.93 \\
   \hline
NB & seas\_growth\_longt\_lag & 10.50 & 10.50 & 0.93 \\
   \hline
\end{tabular}
\caption{Model fit results models without lag transformation.}
\label{damped_no_lag_transformation_mod_info.tex}
\end{table}

\begin{table}[ht]
\centering
\begin{tabular}{|l|l|r|r|r|r|}
  \hline
Family & Parameter Complexity & Percent Damping & RMSE & MAE & cor(obs, sim) \\
  \hline
Poisson & seas\_only & 9.73 & 252.00 & 168.00 & 0.90 \\
   \hline
Poisson & seas\_growth & 12.20 & 177.00 & 110.00 & 0.91 \\
   \hline
Poisson & seas\_growth\_longt & 51.10 & 373.00 & 255.00 & 0.85 \\
   \hline
Poisson & seas\_growth\_lag & 25.20 & 256.00 & 160.00 & 0.84 \\
   \hline
Poisson & seas\_growth\_longt\_lag & 44.70 & 354.00 & 237.00 & 0.85 \\
   \hline
NB & seas\_only & 10.40 & 291.00 & 188.00 & 0.74 \\
   \hline
NB & seas\_growth & 16.60 & 241.00 & 155.00 & 0.81 \\
   \hline
NB & seas\_growth\_longt & 50.10 & 374.00 & 255.00 & 0.82 \\
   \hline
NB & seas\_growth\_lag & 30.90 & 306.00 & 201.00 & 0.79 \\
   \hline
NB & seas\_growth\_longt\_lag & 51.30 & 393.00 & 267.00 & 0.82 \\
   \hline
\end{tabular}
\caption{Simulation results for damped models, without lag transformation}
\label{damped_no_lag_transformation_sim_info.tex}
\end{table}

\subsection{Simulation with lag transformation and no damping}
In this subsection, we perform a simulation with lag transformation (as it is the case
in the main text) but without damping. The simulation results are given in Table
\ref{\labelthree_sim_info.tex}. We observe that the RMSE and MAE in this case are
larger than the corresponding RMSE and MAE in Table \ref{\labelone_sim_info.tex}
which used damping.

\begin{table}[ht]
\centering
\begin{tabular}{|l|l|r|r|r|r|}
  \hline
Family & Parameter Complexity & Percent Damping & RMSE & MAE & cor(obs, sim) \\
  \hline
Poisson & seas\_only & 0.00 & 252.00 & 167.00 & 0.90 \\
   \hline
Poisson & seas\_growth & 0.00 & 178.00 & 110.00 & 0.91 \\
   \hline
Poisson & seas\_growth\_longt & 0.00 & 180.00 & 111.00 & 0.90 \\
   \hline
Poisson & seas\_growth\_lag & 0.00 & 198.00 & 121.00 & 0.88 \\
   \hline
Poisson & seas\_growth\_longt\_lag & 0.00 & 190.00 & 117.00 & 0.88 \\
   \hline
NB & seas\_only & 0.00 & 292.00 & 188.00 & 0.74 \\
   \hline
NB & seas\_growth & 0.00 & 283.00 & 173.00 & 0.78 \\
   \hline
NB & seas\_growth\_longt & 0.00 & 260.00 & 158.00 & 0.80 \\
   \hline
NB & seas\_growth\_lag & 0.00 & 252.00 & 155.00 & 0.80 \\
   \hline
NB & seas\_growth\_longt\_lag & 0.00 & 248.00 & 152.00 & 0.80 \\
   \hline
\end{tabular}
\caption{Simulation results for undamped models with lag transformation.}
\label{undamped_with_lag_transformation_sim_info.tex}
\end{table}







\end{document}